\documentstyle[12pt]{article}
\renewcommand
\baselinestretch{1.4}

\def\rlnabla{\nabla\hspace{-0.9em}\raisebox{1.75ex}{
$\leftrightarrow$}\!\!}

\begin{document}

\hfill HUPD-93-15

\hfill June 1993

\begin{center}

{\LARGE \bf
Covariant effective action and one-loop renormalization of 2D
dilaton gravity with fermionic matter}
\vspace{4mm}

\renewcommand
\baselinestretch{0.8}

{\sc E. Elizalde}\footnote{On leave of absence from and
permanent address: Department E.C.M., Faculty of Physics,
Barcelona University, Diagonal 647, 08028 Barcelona, Spain;
e-mail:
eli
@ ebubecm1.bitnet}
\\ {\it Division of Applied Mechanics, The Norwegian Institute
of Technology, \\ and Department of Theoretical Physics,
University of Trondheim, \\ N-7034 Trondheim, Norway} \\
{\sc S. Naftulin} \\
{\it Institute for Single Crystals, 60 Lenin Ave., 310141
Kharkov,
Ukraine} \\
{\sc S.D. Odintsov}\footnote{ E-mail:
odintsov @ theo.phys.sci.hiroshima-u.ac.jp} \\ {\it
Tomsk Pedagogical Institute, 634041 Tomsk, Russia,} \\ and
{\it
Department of
Physics, Faculty of Science, Hiroshima University, \\
Higashi-Hiroshima 724, Japan}
\medskip

\renewcommand
\baselinestretch{1.3}

\vspace{2mm}

{\bf Abstract}
\end{center}

Two dimensional dilaton gravity interacting with a
four-fermion model and scalars is investigated, all the
 coefficients of the Lagrangian being arbitrary functions
 of the dilaton field. The one-loop covariant
effective action for 2D dilaton gravity with Majorana
spinors (including the four-fermion interaction) is
obtained, and the technical problems which appear in an
attempt at generalizing such calculations to the case
of the most general four-fermion model described by Dirac
fermions are discussed. A solution to these problems is
 found, based on its reduction to the Majorana spinor case.
 The general covariant effective action for 2D dilaton
 gravity with the four-fermion model described
by Dirac spinors is given. The one-loop renormalization of
dilaton gravity with Majorana spinors is carried out and
 the specific conditions for multiplicative renormalizability
are found. A comparison with the same theory but with a
 classical gravitational field is done.

\vspace{1mm}


\newpage

\section{Introduction}

There is an increasing interest in the study of
two-dimensional (2D)
dilaton gravity for different reasons. First, the
insurmountable
difficulties involved in dealing with 4D quantum gravity make
of 2D
dilaton gravity a very interesting laboratory, which may
presumably
lead to the understanding of general properties of true
quantum
gravity. In fact, it is much easier to work with 2D gravity,
because there the by now well-known methods of conformal field
theory can be successfully applied. Second, 2D dilaton gravity
with
matter may well serve as a good toy model to study very
important
features of black hole evaporation and thereby connected
issues
(see [1] for a review and list of references).

Different approaches to the quantization of 2D dilaton
gravities
(mainly, string inspired models) have been discussed in papers
[2-7]  (see also the references therein). Specifically, the
covariant effective action and the one-loop renormalization of
some
specific models have been studied in refs. [3,5,7].

In the present paper we shall obtain the covariant effective
action
corresponding to a very general, multiplicatively
renormalizable
[3,4] (in generalized sense)
 model of 2D gravity with matter. Its action has the following
form
\begin{eqnarray}
S &=& - \int d^2x \sqrt{g} \left[ \frac{1}{2} Z( \Phi) g^{\mu
\nu}
\partial_\mu \Phi \partial_\nu \Phi + C (\Phi ) R- \frac{i}{2}
q (
\Phi) \bar{\psi}_a \gamma^\lambda \partial_\lambda
\psi_a\right.
\nonumber \\
&& \left. + b(\Phi) \left(  \bar{\psi}_a  N_{ab} \psi_b
\right)^2
-\frac{1}{2} f (\Phi) g^{\mu\nu} \partial_\mu \chi_i
\partial_\nu
\chi_i + V(\Phi, \chi) \right].
\end{eqnarray}
It includes a dilaton field, $\Phi$, $n$ Majorana fermions
$\psi_a$
interacting quartically via a symmetric constant matrix
$N_{ab}$,
and $m$ real scalars $\chi_i$. We shall also consider the
(much
more difficult) case in which this action contains 2D Dirac
fermions. Notice that we have chosen the matter to interact
with
the dilaton via arbitrary functions.

This action describes and generalizes many well-known dilaton
models. For instance, the celebrated bosonic string
effective action corresponds to
\begin{equation}
Z(\Phi )= 8 e^{-2\Phi}, \ C(\Phi) =  e^{-2\Phi}, \ V(\Phi)  =4
\lambda^2  e^{-2\Phi}, \ q(\Phi) =  b(\Phi) = 0, \  f(\Phi)
=1.
\end{equation}
On the other hand, in the absence of matter our action for
\begin{equation}
Z=0, \ \ \ \ C(\Phi) =  \Phi, \ \ \ \ V(\Phi)  = \Lambda \Phi,
\end{equation}
coincides with the Jackiw-Teitelboim action [8]. But other
dilaton
models are parametrized by the set $\left\{ Z, C, q, b, f, V
\right\}$. It goes without saying that one can also add gauge
fields to the matter sector.

In principle, at the classical level the theory defined by the
action (1) can be transformed into an equivalent theory whose
corresponding action is  more simple. Indeed, this  can be
done
by choosing the field $\varphi_1$ as defined through the
equation
\begin{equation}
Z^{1/2} (\Phi) \partial_\mu \Phi = \partial_\mu \varphi_1,
\end{equation}
and by expressing $\Phi$ as $\Phi = \Phi (\varphi_1) $. Next,
let us
introduce a new field, $\varphi_2$, via
\begin{equation}
c \varphi_2 = C ( \Phi (\varphi_1) ),
\end{equation}
and write then  $\Phi = \Phi (\varphi_2) $. After having done
this, by making the transformation [6]
\begin{equation}
g_{\mu\nu} \longrightarrow e^{2 \rho (\varphi_2)}
\bar{g}_{\mu\nu},
\end{equation}
with a properly chosen $ \rho (\varphi_2)$ [6],  one can see
that
the theory (1) with the transformed metric (6) is classically
equivalent to the more simple, particular case
\begin{eqnarray}
&& Z=1, \ \ C= c\varphi_2, \ \ V= e^{2 \rho (\varphi_2)} V(
\varphi_2, \chi ), \ \ f (\Phi ) = f(\varphi_2), \nonumber \\
&& b (\Phi) = b (\varphi_2) \, e^{2 \rho (\varphi_2)}, \ \ q
(\Phi)
= q (\varphi_2) \, e^{\rho (\varphi_2)}.
\end{eqnarray}
However, the model (7) (which, generally speaking, may be
considered as a representative of the general class (1)) is
still
complicated enough. Moreover, it still includes arbitrary
functions
of the dilaton (now $\varphi_2$), as (1) does. Finally, the
classical equivalence may be lost at the quantum level. For
all these
reasons, we choose to consider the quantum effective action
corresponding
to the more general theory (1).

Being more specific, in this paper we shall construct the
covariant
effective action of the theory (1), study its one-loop
renormalization and discuss some thereby connected issues. The
work
is organized as follows. In the next section we describe  in
full
detail the calculation of the one-loop covariant effective
action
in 2D dilaton gravity with Majorana spinors. This is, to our
knowledge, the first example of such a kind of calculation in
two
dimensions. The inclusion of scalars is also discussed in that
section.  Sect. 3 is devoted to the computation of the
covariant
effective action of dilaton, scalars and Majorana spinors for
quantum systems in classical spacetime. In Sect. 4 we discuss
the
technical problems which appear in the derivation of the
covariant
effective action in 2D dilaton gravity with Dirac fermions.
The
solution of these problems is found, via reduction of the
system to
the previously discussed case of the theory of quantum dilaton
gravity with Majorana spinors. In Sect. 5 the one-loop
renormalization of quantum dilaton gravity with Majorana
spinors is
discussed. The conditions of multiplicative renormalizability
are
specified and some examples of multiplicatively renormalizable
dilaton potentials are obtained. Finally, the conclusions of
the
paper are presented in Sect. 6. There is also a short appendix
on the sine-Gordon model.
\bigskip

\section{The one-loop effective action of 2D quantum gravity
with
matter}

In this section we will calculate the covariant effective
action
for the theory given by the action (1). For simplicity, we
first
put the scalars $\chi_i =0$ and discuss dilaton-Majorana
gravity
only, i.e., the action
\begin{eqnarray}
S &=& - \int d^2x \sqrt{g} \left[ \frac{1}{2} Z( \Phi) g^{\mu
\nu}
\partial_\mu \Phi \partial_\nu \Phi + C (\Phi ) R- \frac{i}{2}
q (
\Phi) \bar{\psi}_a \gamma^\lambda \partial_\lambda \psi_a
\right.
\nonumber \\
&& \left. + b(\Phi) \left(  \bar{\psi}_a  N_{ab} \psi_b
\right)^2
 + V(\Phi) \right].
\end{eqnarray}
Our purpose is to calculate first the effective action for the
theory (8), before proceding with the general case.

Let us introduce some notations:
\begin{equation}
T^{\mu\nu} = - \frac{i}{4}  \left(
 \bar{\psi}_a \gamma^\mu \nabla^\nu  \psi_a + \bar{\psi}_a
\gamma^\nu \nabla^\mu  \psi_a \right), \ \ T = T^\mu_{\ \mu},
\ \
J =  \bar{\psi}_a  N_{ab} \psi_b.
\end{equation}
Notice that $T_{\mu \nu}$ is {\it not} the stress tensor. We
are
going to use the formalism of the background field method,
representing
\begin{equation}
\psi \longrightarrow \psi + \eta, \ \ \ \ \Phi \longrightarrow
\Phi
+ \varphi, \ \ \ \ e^\mu_a \longrightarrow  e^\mu_a +  h^\mu_a
,
\end{equation}
where $\psi$, $\Phi$ and  $ e^\mu_a$ are background fields.
The
action (8) will be expanded to second order in quantum fields.
 The Lorentz symmetry of the zweibein $(e^\mu_a)$ is
`frustrated' by imposing the no-torsion condition
$e_{a\mu} h^\mu_b- e_{b\mu} h^\mu_a=0$ as a constraint: we
insert
the
corresponding delta-function into the path integral but do not
exponentiate
it. The corresponding ghost contribution, which is
proportional to
$\delta(0)$,
may be discarded. This procedure fixes one out of four gauge
parameters and leaves
three to be undone by the metric variations. Thus, the
variation
of the zweibein
$h^\mu_a$ is solved for the metric variation $h_{\mu\nu}\,$
\begin{equation}
h^\mu_a={1\over2}h^\mu_\lambda
e^\lambda_a+{3\over8}h^{\mu\lambda}
        h_{\lambda\nu}e^\nu_a+ \dots.
\end{equation}
That is,
 we will work with the variations of the metric rather than
the
zweibein.
The quantum fields are arranged into a vector
$\phi^i=\{\varphi;h;\bar{h}_{\mu\nu};\eta_a\}$, where
$\bar{h}_{\mu\nu}
= h_{\mu\nu} - \frac{1}{2} g_{\mu\nu}$.
 The equations of motion with respect to the background fields
are
\begin{eqnarray}
&&
-{\delta_{\vphantom{p}_R}S\over\delta\overline{\psi}}=0
=iq\gamma^\lambda
\nabla_\lambda\psi
+i{q'\over2}(\nabla_\lambda\Phi)(\gamma^\lambda
\psi)-
    4bJ(N\psi) \ ,   \\ \cr
&&  -{\delta S\over
\delta\Phi}=0=-Z(\Delta\Phi)-{1\over2}Z'(\nabla^\lambda
    \Phi)(\nabla_\lambda\Phi)+C'R+V'+qT+bJ^2 \ ,   \\ \cr
&&  -g_{\mu\nu}{\delta S\over\delta g_{\mu\nu}}=0=-\Delta
C+V+{1\over2}qT+ bJ^2 \ .
\end{eqnarray}
Here $\delta_{\vphantom{p}_R}$ denotes the right functional
derivative.

The covariant gauge fixing condition is
\begin{eqnarray}
&  S_{g.f.}=-{1\over2}\int c_{\mu\nu}\chi^\mu\chi^\nu \ ,  &
\cr\cr
&  \chi^\mu =-\nabla_\nu\bar{h}^{\mu\nu}+{C'\over
C}\nabla^\mu\varphi+
   {iq\over4C}\overline{\psi}\gamma^\mu\eta \ , \qquad
   c_{\mu\nu}=-C\sqrt{g}\,g_{\mu\nu} \ ,   &
\label{gauge_fix}
\end{eqnarray}
and the total quadratic contribution to the action takes the
form
$$
S^{(2)}_{tot}=-{1\over2}\int
d^2x\,\sqrt{g}\phi^i\widehat{H}_{ij}\phi^j \ ,
$$
where $\widehat{H}$ is the second order minimal operator
(though the
coefficient matrix multiplying the Laplacian is not
invertible,
since the fermionic fields are of first order). Following ref.
[9], we
introduce another operator
\begin{equation}
\widehat{\Omega}_{ij}=\pmatrix{1 & 0 & 0 & 0 \cr 0 & 1 & 0 & 0
\cr
                               0 & 0 & P_{\mu\nu,\mu'\nu'} & 0
\cr
                               0 & 0 & 0 &
-i\delta_{ab}\gamma^\sigma\nabla_\sigma}
\end{equation}
and define $\widehat{\cal H}=\widehat{H}\widehat{\Omega}$ so
that
the one-loop
effective action in terms of supertraces becomes
\begin{equation}
\Gamma={i\over2}\mbox{Tr}\,\log\widehat{\cal H}
-{i\over4}\mbox{Tr}\,\log\widehat{\Omega}^2
        -i\mbox{Tr}\,\log\widehat{\cal M} \ ,
\label{def_eff_action}
\end{equation}
where, as usual, $\widehat{\cal
H}=-\widehat{K}\Delta+\widehat{L}^\lambda
\nabla_\lambda+\widehat{P}$, and the last term in
(\ref{def_eff_action}) is
the ghost operator corresponding via (\ref{gauge_fix}) to
diffeomorphisms.

Now,
\begin{equation}
\widehat{K}_{ij}=\pmatrix{Z-{{C'}^2\over C} & C'/2 & 0 &
q'\overline{\psi}\cr
                      C'/2 & 0 & 0 & (q/4)\overline{\psi}\cr
                      0 & 0 & -{C\over2}P^{\mu\nu,\alpha\beta}
&
0\cr
                      0 & 0 & 0 & q\hat1 \cr}
\end{equation}
and the other (essential) matrix elements%
\footnote{In the non-spinor sectors the 't Hooft \& Veltman
doubling
procedure [10] is assumed.}
are:
\begin{eqnarray}
&&\widehat{L}^\lambda_{11}=\left(2{C'C''\over C}-{{C'}^3\over
C^2}-Z'\right)
                      (\nabla^\lambda\Phi) \ ,  \cr\cr
&&\widehat{L}^\lambda_{12}=0 \ ,\cr\cr
&&\widehat{L}^\lambda_{13}=-\widehat{L}^\lambda_{31}
=(Z-C'')(\nabla_\omega
                            \Phi)P^{\alpha\beta,\lambda\omega
} \
,   \cr\cr
&&\widehat{L}^\lambda_{14}=0 \ , \cr\cr
&&\widehat{L}^\lambda_{21}=-C''(\nabla^\lambda\Phi) \ , \cr\cr
&&\widehat{L}^\lambda_{22}=0 \ ,\cr\cr
&&\widehat{L}^\lambda_{23}=-\widehat{L}^\lambda_{32}={1\over2}
C'(
\nabla_\omega\Phi)P^{\alpha\beta,\lambda\omega}\ ,\cr\cr
&&\widehat{L}^\lambda_{24}=0 \ , \cr\cr
&&\widehat{L}^\lambda_{33}=C'(\nabla^\omega\Phi)
\left[P^{\mu\nu}_{\omega
\kappa}P^{\alpha\beta,\lambda\kappa}-P^{\mu\nu,
\lambda\kappa}P^{\alpha\beta}_{\omega\kappa}\right]+{1\over2}C
'
         (\nabla^\lambda\Phi)P^{\mu\nu,\alpha\beta} \ ,
\cr\cr
&&\widehat{L}^\lambda_{34}=-{1\over2}q'(\nabla^\omega\Phi)
(\overline{\psi}
\gamma^\kappa\gamma^\lambda)P^{\mu\nu}_{\omega
                            \kappa} \ ,    \cr\cr
&&\widehat{L}^\lambda_{41}=i{qC'\over2C}(\gamma^\lambda\psi) \
,
\cr\cr
&&\widehat{L}^\lambda_{42}=\widehat{L}^\lambda_{43}=0 \ ,
\cr\cr
&&\widehat{L}^\lambda_{44}=-4ibJN\gamma^\lambda-8ib(N\psi)
(\overline{\psi}N)
\gamma^\lambda+i{q^2\over16C}(\gamma_\kappa\psi)
         (\overline{\psi}\gamma^\kappa\gamma^\lambda) \ ,
\cr\cr
&&\widehat{P}_{12}={1\over2}V'+{1\over4}q'T+{1\over2}b'J^2 \ ,
\cr\cr
&&\widehat{P}_{21}=-{1\over2}C''(\Delta\Phi)-{1\over2}C'''
(\nabla^\lambda\Phi)
(\nabla_\lambda\Phi)+{1\over2}V'+{1\over4}q'T+{1\over2}b'
                   J^2 \ ,     \cr\cr
&&\widehat{P}_{22}={1\over4}C'(\Delta\Phi)+{1\over4}C''
(\nabla^\lambda\Phi)
               (\nabla_\lambda\Phi)-{1\over 16} qT \ ,
\cr\cr
&&\widehat{P}_{33}
=\Bigl[\bigl(Z-2C''\bigr)(\nabla_\omega\Phi)(\nabla^\lambda
\Phi)-2C'(\nabla_\omega\nabla^\lambda\Phi)+
{3\over4}qT_\omega^\lambda\Bigr]P^{\mu\nu,\omega\kappa}
P^{\alpha\beta}_{\lambda
         \kappa}    \cr\cr
&&\phantom{\widehat{P}_{33}=}
-{1\over2}\left[CR+V+qT+bJ^2+({1\over2}Z-2C'')
(\nabla^\lambda\Phi)(\nabla_\lambda\Phi)-2C'
         (\Delta\Phi)\right]P^{\mu\nu,\alpha\beta} \ ,
\cr\cr
&&\widehat{P}_{41}=8b'J(N\psi)-iq'(\gamma^\lambda
\nabla_\lambda\psi)\
 , \cr\cr
&&\widehat{P}_{42}=4bJ(N\psi)-{i\over4}q(\gamma^\lambda
\nabla_\lambda\psi) \ ,
  \cr\cr
&&\widehat{P}_{44}={1\over4}qR\hat1 \ .
\end{eqnarray}
To calculate the divergent part we use the same technique as
in ref. [3],
representing
\begin{equation}
 \widehat{\cal H}=- \widehat{K} \left( \widehat{1} \Delta +
2 \widehat{E}^\lambda \nabla_\lambda+ \widehat{\Pi} \right), \
\ \
\
 \widehat{E}^\lambda =-\frac{1}{2} \widehat{K}^{-1}
 \widehat{L}^\lambda, \ \ \ \
 \widehat{\Pi} =-  \widehat{K}^{-1}  \widehat{P}.
\end{equation}
After that, the standard algorithm can be use, namely
\begin{equation}
\frac{i}{2} \mbox{Tr}\, \log  \widehat{\cal H}=
\left. \frac{i}{2} \mbox{Tr}\, \log  \left( \widehat{1} \Delta
+ 2 \widehat{E}^\lambda \nabla_\lambda+ \widehat{\Pi}
\right)\right|_{div} = \frac{1}{2\epsilon} \int d^2 x \,
\sqrt{g} \,
\mbox{Tr}\, \left( \widehat{\Pi}+ \frac{R}{6} \widehat{1}
- \widehat{E}^\lambda  \widehat{E}_\lambda - \nabla_\lambda
 \widehat{E}^\lambda \right),
\end{equation}
where $\epsilon = 2\pi (n-2)$.
The matrices $\hat{E}^\lambda$ and $\hat{\Pi}$ are given by:
\begin{eqnarray}
&&\big(\hat{E}^\lambda\big)^{1}_{1}={C''\over
C'}(\nabla^\lambda\Phi) \
,  \cr\cr
&&\big(\hat{E}^\lambda\big)^{1}_{2}=0 \ ,     \cr\cr
&&\big(\hat{E}^\lambda\big)^{1}_{3}=-{1\over
2}(\nabla_\omega\Phi)P^{\alpha
                                    \beta,\lambda\omega} \ ,
\cr\cr
&&\big(\hat{E}^\lambda\big)^{1}_{4}=-3i{b\over
C'}J(\overline{\psi}N\gamma^\lambda) \ ,    \cr\cr
&&\big(\hat{E}^\lambda\big)^{2}_{1}=\left({Z'\over
C'}-2{C''Z\over{C'}^2}+
                                    {{C'}^2\over
C^2}\right)(\nabla^\lambda
                                    \Phi) \ ,     \cr\cr
&&\big(\hat{E}^\lambda\big)^{2}_{2}=0 \ ,     \cr\cr
&&\big(\hat{E}^\lambda\big)^{2}_{3}=\left({C''\over
C'}-{C'\over
C}\right)
(\nabla_\omega\Phi)P^{\alpha\beta,\lambda
                                    \omega} \ ,     \cr\cr
&&\big(\hat{E}^\lambda\big)^{2}_{4}=6ibJ\left({Z\over{C'}^2}-
{1\over C}- 2{q'\over
qC'}\right)(\overline{\psi}N \gamma^\lambda) \ ,     \cr\cr
&&\big(\hat{E}^\lambda\big)^{3}_{1}=\left({C''\over C}-{Z\over
C}\right)
(\nabla_\omega\Phi)
P^{\lambda\omega}_{\rho\sigma} \ ,   \cr\cr
&&\big(\hat{E}^\lambda\big)^{3}_{2}=-{C'\over2C}
(\nabla_\omega\Phi)P^{\lambda
                                    \omega}_{\rho\sigma} \ ,
\cr\cr
&&\big(\hat{E}^\lambda\big)^{3}_{3}={C'\over
C}(\nabla^\omega\Phi)\Bigl[P_{\rho\sigma,\omega\kappa}
P^{\alpha\beta,
\lambda\kappa}-P^{\alpha\beta}_{\omega\kappa}
P^{\lambda\kappa}_{\rho
\sigma}\Bigr]+{C'\over2C}(\nabla^\lambda\Phi)
P^{\alpha\beta}_{\rho
         \sigma}  \ ,          \cr\cr
&&\big(\hat{E}^\lambda\big)^{3}_{4}=-{q'\over2C}(\nabla^\omega
\Phi)
(\overline{\psi}\gamma^\kappa
\gamma^\lambda)P_{\rho\sigma,\omega\kappa} \ ,     \cr\cr
&&\big(\hat{E}^\lambda\big)^{4}_{1}
=-i{C'\over4C}(\gamma^\lambda\psi), \cr\cr
&&\big(\hat{E}^\lambda\big)^{4}_{2}
=\big(\hat{E}^\lambda\big)^{4}_{3}=0
\ ,
   \cr\cr
&&\big(\hat{E}^\lambda\big)^{4}_{4}=2i{b\over
q}J(N\gamma^\lambda)+4i{b\over
q}(N\psi)(\overline{\psi}N\gamma^\lambda)
-i{q\over32C}(\gamma_\nu\psi)
(\overline{\psi}\gamma^\nu\gamma^\lambda) \ ,     \cr\cr
&&\hat{\Pi}^{1}_{1}={C'''\over
C'}(\nabla^\lambda\Phi)(\nabla_\lambda\Phi)+
                    {C''\over C'}(\Delta\Phi)-{1\over
C'}V'+{q'\over2C'}T+
                    3{b'\over C'}J^2 \ ,    \cr\cr
&&\hat{\Pi}^{2}_{2}=\left({C''Z\over{C'}^2}-{C''\over
C}\right)(\nabla^\lambda
                    \Phi)(\nabla_\lambda\Phi)+\left({Z\over
C'}-{C'\over C}
         \right)(\Delta\Phi)-{1\over C'}V'   \cr\cr
&&\phantom{\hat{\Pi}^{3}_{3}=}
+\left({3q\over4C}+{q'\over2C'}-{3Zq'\over4
                  {C'}^2}\right)T+\left({4b\over
C}+{8bq'\over
         qC'}-{4bZ\over{C'}^2}-{b'\over C'}\right)J^2 \ ,
\cr\cr
&&\hat{\Pi}^{3}_{3}=\biggl[\left({2Z\over C}-{4C''\over
C}\right)(\nabla_\omega\Phi)(\nabla^\lambda\Phi)
-{4C'\over
C}(\nabla_\omega\nabla^\lambda\Phi)
+{3q\over2C}T_\omega^\lambda\biggr]P^{\omega
         \kappa}_{\rho\sigma}P^{\alpha\beta}_{\lambda\kappa}
\cr\cr
&&\phantom{\hat{\Pi}^{3}_{3}=} +\biggl[\left({2C''\over
C}-{Z\over
2C}\right)
(\nabla^\lambda\Phi)(\nabla_\lambda\Phi)+2{C'
         \over C}(\Delta\Phi)-R-{1\over C}V-{q\over
C}T-{b\over
C}J^2\biggr]
         P^{\alpha\beta}_{\rho\sigma} \ ,     \cr\cr
&&\hat{\Pi}^{4}_{4}=-{1\over4}R\hat1 \ .
\end{eqnarray}
To obtain the divergent part, $\Gamma_{2-div}$, we have to
evaluate
the
functional (super)traces of the matrices above according to
(21).
After
some tedious algebra we arrive at
\begin{eqnarray}
\Gamma_{2-div} =\frac{i}{2} \mbox{Tr}\, \log  \widehat{\cal H}
=-{1\over 2\epsilon}\int d^2x\,\sqrt{g}\,\Biggl\{{8-n\over6}R
                 +{2\over C}V+{2\over C'}V'+\left({2C'\over
C}-{Z\over C'}
        \right)(\Delta\Phi)  \qquad\qquad\cr\cr
-\left({3{C'}^2\over2C^2}+{C''Z\over{C'}^2}\right)
(\nabla^\lambda
\Phi)(\nabla_\lambda\Phi)+\left({3qZ\over4{C'}^2}-{q\over4C}-{
q'\over
C'}\right)T
                             \qquad\qquad\cr\cr
+\left({4bZ\over{C'}^2}-{4b\over C}-{2b'\over C'}-{8bq'\over
qC'}+{16b^2\over
         q^2}(N_{ab}N_{ba})\right)J^2-{32b^2\over
q^2}J(\overline{\psi}N^3
         \psi)\Biggr\},
\end{eqnarray}
where we have used the Majorana identity
$\overline{\psi}\gamma^\lambda
N^k\psi=0$ for all
$k$.

The next step is to subtract the squared contribution, i.e.
the
second term
in Eq. (\ref{def_eff_action}):
\begin{equation}
-\left. \frac{i}{4}
\mbox{Tr}\,\log\widehat{\Omega}^2\right|_{div}
=\left.\frac{i}{4} \mbox{Tr}\,\log\left(-\Delta+{R\over 4}
                                       \hat1
\right)\right|_{div}=
         -{1\over 2\epsilon}\int d^2x\, \sqrt{g} \,  {n\over
12}R \ .
\end{equation}

To complete the calculation one has to consider the ghost
operator
$$
\widehat{\cal
M}^\mu_\nu\equiv{\delta_{\vphantom{p}_R}\chi^\mu\over\delta
                              \phi^j}\nabla^j_\nu \ ,
$$
where again $\delta_{\vphantom{p}_R}$ stands for the right
functional
derivative and the generators of the diffeomorphisms are
\begin{equation}
\nabla^\varphi_\nu=-(\nabla_\nu \Phi) \ , \qquad
\nabla^{\bar{h}_{\rho\sigma}}_\nu=g_{\rho\sigma}\nabla_\nu
-g_{\nu
\rho}\nabla_\sigma-g_{\nu\sigma}\nabla_\rho
\ , \qquad
\nabla^\eta_\nu=
-(\nabla_\nu\psi)+{1\over8}\left[\gamma_\nu,\gamma_\lambda
                \right]\psi\nabla^\lambda \ .
\end{equation}
Explicitly,
\begin{equation}
\widehat{\cal M}^\mu_\nu=g^\mu_\nu\Delta-{C'\over
C}(\nabla_\nu\Phi)\nabla^\mu-{C'\over
C}(\nabla^\mu\nabla_\nu\Phi)+
R^\mu_\nu-i{q\over4C}(\overline{\psi}\gamma^\mu\nabla_\nu\psi)
\ ,
\end{equation}
which leads to
\begin{equation}
\Gamma_{gh-div}=-{1\over 2\epsilon}\int
d^2x\,\sqrt{g}\,\Biggl\{{8\over3}R
                -{C'\over C}(\Delta\Phi)+\left({C''\over
C}-{3{C'}^2\over
2C^2}\right)(\nabla^\lambda\Phi)(\nabla_\lambda\Phi)+
                {q\over C}T\Biggr\} \ .
\end{equation}
The final answer is
\begin{eqnarray}
\Gamma_{div}=-{1\over 2\epsilon}\int
d^2x\,\sqrt{g}\,\Biggl\{{48-n\over 12}R
             +{2\over C}V+{2\over C'}V'+\left({C'\over
C}-{Z\over
C'}\right)
             (\Delta\Phi)      \qquad\qquad  \cr\cr
+\left({C''\over C}-3{{C'}^2\over
C^2}-{C''Z\over{C'}^2}\right)(\nabla^\lambda
              \Phi
)(\nabla_\lambda\Phi)+\left({3qZ\over4{C'}^2}+{3q\over4C}-
              {q'\over C'}\right)T     \qquad\qquad  \cr\cr
+\left({4bZ\over{C'}^2}-{4b\over C}-{2b'\over C'}-{8bq'\over
qC'}+{16b^2\over
         q^2}(N_{ab}N_{ba})\right)J^2-{32b^2\over
q^2}J(\overline{\psi}N^3
         \psi)\Biggr\} \ .
\end{eqnarray}
\smallskip
Notice that all surface terms have been kept in Eq. (28).

Having performed the above calculation, now it is not
difficult to
take
into account the scalar fields $\chi_i$, and to repeat it for
the
effective action (1). In fact, the introduction of the scalars
leads to minor changes. All matrices become $5\times 5$
(rather
than $4\times 4$): in the background field notation $\chi_i
\rightarrow \chi_i + \sigma_i$ the quantum fields are $\phi^i
= \{
\varphi; h; \bar{h}_{\mu\nu}; \eta_a; \sigma_i\}$. Scalars do
not
spoil the minimality of the second functional derivative
operator,
so that the gauge condition may be left untouched. As a
consequence, neither ghost terms nor the squared contributions
change, and only a few matrix elements of
$\widehat{L}^\lambda$ and
$\widehat{P}$ do. Besides the completely new elements:
\begin{eqnarray}
&&\widehat{L}^\lambda_{15}=-\widehat{L}^\lambda_{51}=-
f'(\nabla^\lambda\chi) \ ,  \cr\cr
&&\widehat{L}^\lambda_{25}=\widehat{L}^\lambda_{52}=
\widehat{L}^\lambda_{45}=\widehat{L}^\lambda_{54}=0 \ ,
\cr\cr
&&\widehat{L}^\lambda_{35}=-\widehat{L}^\lambda_{53}=
f(\nabla_\omega\chi) P^{\mu\nu,\lambda\omega} \ ,  \cr\cr
&&\widehat{L}^\lambda_{55}= f'(\nabla^\lambda\Phi) \hat1 \ ,
\cr\cr
&&\widehat{P}_{55} = V_{,ij},
\end{eqnarray}
there is only one extra contribution
\begin{equation}
\widehat{P}_{33 \ extra \ terms} = \frac{1}{4}
f(\nabla^\lambda\chi) (\nabla_\lambda\chi)
P^{\mu\nu,\alpha\beta}
-f(\nabla_\omega\chi) (\nabla^\lambda\chi)
P^{\mu\nu,\omega\kappa}
P^{\alpha\beta}_{\lambda\kappa},
\end{equation}
which is essential for finding the divergences. The 't Hooft
\&
Veltman procedure [10] is assumed to have been implemented
whenever
necessary, and we use the notations \[
V' \equiv \frac{\delta V}{\delta \Phi} , \ \ \ \ V_{,i} \equiv
\frac{\delta V}{\delta \chi_i}.
\]

The matrix $\widehat{K}$ becomes
\begin{equation}
\widehat{K} = \left( \begin{array}{ccccc} Z-{C'}^2/C & C'/2 &
0 & q'
\bar{\psi} & 0 \\
C'/2 & 0 & 0 &  (q/4) \bar{\psi} & 0 \\
0 & 0 & -(C/2)  P^{\mu\nu,\alpha\beta}  & 0 & 0 \\
0 & 0 & 0 & q \hat1 & 0 \\
0 & 0 & 0 & 0 & -f \hat1 \end{array} \right).
\end{equation}
Repeating the above procedure we may calculate the extra terms
which appear in the effective action as a result of the scalar
fields contribution:
\begin{eqnarray}
\Gamma_{2-div \ extra \ terms}& =&-{1\over2\epsilon}\int
d^2x\,\sqrt{g}\,\left\{
-{m\over6}R- \frac{1}{f} V_{,ii}+ \frac{mf'}{2f} \Delta \Phi
\right. \nonumber \\
&& + \left. \left( \frac{mf''}{2f}- \frac{m{f'}^2}{4f^2}
\right)
(\nabla^\lambda \Phi)(\nabla_\lambda \Phi) \right\}.
\end{eqnarray}
Thus, the complete one-loop divergences for the theory (1)
become
\begin{eqnarray}
\Gamma_{div}=-{1\over2\epsilon}\int d^2x\,\sqrt{g}\,\Biggl\{
 {48-n+2m\over12}R+{2\over C} V+ {2\over C'} V' -
{V_{,ii}\over f}  \qquad\qquad  \cr\cr
+ \left( {C''\over C}-{3{C'}^2\over C^2}- {C''Z\over {C'}^2} -
{m{f'}^2\over4f^2}+ {mf''\over2f} \right)(\nabla^\lambda\Phi
)(\nabla_\lambda\Phi) \qquad\qquad  \cr\cr
+ \left( {C'\over C}- {Z\over
C'}+{mf'\over2f}\right)\Delta\Phi
+\left({3qZ\over4{C'}^2}+{3q\over 4C}-{q'\over C'}\right)T
\qquad\qquad  \cr\cr
+\left({4bZ\over {C'}^2}-{4b\over C}-{2b'\over C'}-
{8bq'\over qC'}+{16b^2\over
q^2}(N_{ab}N_{ba})\right)
         J^2-{32b^2\over q^2}J(\overline{\psi}N^3\psi)\Biggr\}
\ .
\end{eqnarray}
This is the main result of the present section ---the one-loop
effective action for 2D dilaton
gravity with matter. Furthermore, one can
easily generalize this expression to the case when Maxwell
fields
are added, namely, when one considers action (1) plus the
Maxwell
action:
\begin{equation}
S= -\int d^2x\,\sqrt{g}\,\left[ \cdots + \frac{1}{4} f_1
(\Phi)
F^2_{\mu\nu} \right]. \nonumber
\end{equation}
With the background vs. quantum field separation
$A_\mu\rightarrow
A_\mu +Q_\mu$, in the Lorentz gauge,
\begin{equation}
S_{Lorentz}= -\int d^2x\,\sqrt{g}\, f_1 (\Phi) (\nabla_\mu
Q_\mu)^2,
\end{equation}
the extra contributions are known to split into the
$F^2$-terms and
total divergences (see the second and third refs. [5]), which
may
be calculated independently, to yield
\begin{eqnarray}
\Gamma_{Max.,div}^{extra}&=& -\frac{1}{2\epsilon}\int
d^2x\,\sqrt{g}\,\left[ R + \left(\frac{{f_1}'}{2C'}-
\frac{f_1}{2C}
\right) F^2_{\mu\nu}+\frac{{f_1}'}{f_1} \Delta\Phi \right.
\nonumber
\\
&&+ \left.  \left(\frac{{f_1}''}{f_1}- \frac{{f_1'}^2}{f_1^2}
\right) (\nabla^\lambda\Phi )(\nabla_\lambda\Phi)  \right].
\end{eqnarray}

Thus, the total divergent contribution to the one-loop
effective
action of the theory (1) plus the Maxwell terms (34) is given
by
\begin{eqnarray}
\Gamma_{div}=-{1\over2\epsilon}\int d^2x\,\sqrt{g}\,\Biggl\{
 {-n-2m+60\over12}R+{2\over C} V+ {2\over C'} V' -
{V_{,ii}\over f}  + \left(\frac{{f_1}'}{2C'}- \frac{f_1}{2C}
\right)
F^2_{\mu\nu} \qquad\qquad  \cr\cr
+ \left( {C''\over C}-{3{C'}^2\over C^2}- {C''Z\over {C'}^2} -
{m{f'}^2\over4f^2}+ {mf''\over2f}+\frac{{f_1}''}{f_1}-
\frac{{f_1'}^2}{f_1^2} \right)(\nabla^\lambda\Phi
)(\nabla_\lambda\Phi) \qquad\qquad  \cr\cr
+ \left( {C'\over C}-
{Z\over C'}+{mf'\over2f}+\frac{{f_1}'}{f_1}\right)\Delta\Phi
+\left({3qZ\over4{C'}^2}+{3q\over 4C}-{q'\over C'}\right)T
\qquad\qquad  \cr\cr
+\left({4bC\over {C'}^2}-{4b\over C}-{2b'\over C'}-
{8bq'\over qC'}+{16b^2\over
q^2}(N_{ab}N_{ba})\right)
         J^2-{32b^2\over q^2}J(\overline{\psi}N^3\psi)\Biggr\}
\ .
\end{eqnarray}
This gives the divergences of the covariant effective action
(for
a recent discussion of the covariant effective action
formalism, see
[11], general review is given in [14]).
 The renormalization of the theory (1) using the one-loop
effective action (33) will be discussed in the next sections.
\bigskip

\section{The one-loop effective action in curved spacetime}

In this section we will calculate the one-loop effective
action for
the theory (1) in the case when the gravitational field is a
classical one, but the dilaton and the rest of the matter
fields
are quantized. Such a calculation is much simpler than the one
carried out in the previous section, since there are no
gauge-fixing terms and corresponding ghosts.

The effective action has the form
\begin{equation}
\Gamma={i\over2}\mbox{Tr}\,\log\widehat{\cal H}-\frac{i}{4}
\mbox{Tr}\,\log\widehat{\Omega}^2
\end{equation}
where
\begin{equation}
\widehat{\cal
H}=-\widehat{K}\Delta+\widehat{L}^\lambda\nabla_\lambda+
\widehat{P}, \nonumber
\end{equation}
and the quantum fields are arranged in the vector form $\{
\varphi'; {\eta'}_a; \sigma_i \}$.

The basic matrices which enter in the operator of small
disturbances are
\begin{equation}
\widehat{\Omega}=\pmatrix{1 & 0 & 0 \cr 0 &
-i\gamma^\lambda\nabla_\lambda\hat1 & 0
                          \cr 0 & 0 & \hat1 \cr}\ ,  \qquad
\widehat{K}=\pmatrix{Z & q'\overline{\psi} & 0 \cr 0 & q\hat1
& 0
\cr 0 & 0 & -f \hat1 \cr},
\end{equation}
and the $\widehat{L}^\lambda$ and $\widehat{P}$ can be
calculated easily. The divergent part of the effective action
is expressed in terms of the traces of the following matrices
(see also (20)):
\begin{eqnarray}
&&\big(\hat{E}^\lambda\big)^{1}_{1}={Z'\over2Z}(\nabla^\lambda
\Phi)
               \ ,  \cr\cr
&&\big(\hat{E}^\lambda\big)^{1}_{2}=-6i{bq'\over
qZ}J(\overline{\psi}N
                                     \gamma^\lambda)  \ ,
\cr\cr
&&\big(\hat{E}^\lambda\big)^{1}_{3}={f'\over2Z}(\nabla^\lambda
\chi)
               \ ,  \cr\cr
&&\big(\hat{E}^\lambda\big)^{2}_{1}=0  \ ,  \cr\cr
&&\big(\hat{E}^\lambda\big)^{2}_{2}=2i{b\over
q}JN\gamma^\lambda+4i{b\over q}
(N\psi)(\overline{\psi}N\gamma^\lambda)
                   \ ,  \cr\cr
&&\big(\hat{E}^\lambda\big)^{2}_{3}
=\big(\hat{E}^\lambda\big)^{3}_{2}=0  \ ,  \cr\cr
&&\big(\hat{E}^\lambda\big)^{3}_{1}={f'\over2f}(\nabla^\lambda
\chi)
               \ ,  \cr\cr
&&\big(\hat{E}^\lambda\big)^{3}_{3}={f'\over2f}(\nabla^\lambda
\Phi)
\hat1      \ ,
\end{eqnarray}
and
\begin{eqnarray}
&&\big(\hat{\Pi}\big)^{1}_{1}={Z''\over2Z}(\nabla^\lambda\Phi)
(\nabla_\lambda \Phi)+{Z'\over Z}(\Delta\Phi)-{C''\over Z}R
         -{V''\over Z}+\left({2{q'}^2\over qZ}-{q''\over
Z}\right)T
\cr\cr
&&\phantom{\big(\hat{\Pi}\big)^{1}_{1}=} +\left({8b'q'\over
qZ}-{q''\over Z}
                                         \right)J^2 +
\frac{f''}{2Z} (\nabla^{\lambda} \chi)
(\nabla_{\lambda} \chi)    \ ,  \cr\cr
&&\big(\hat{\Pi}\big)^{2}_{2}=-{1\over4}R\hat1, \cr\cr
&&\big(\hat{\Pi}\big)^{3}_{3}={1\over f}V_{,ij}  \ .
\end{eqnarray}
We get
\begin{equation}
\Gamma_{div}=-{1\over2\epsilon}\int d^2x\,\sqrt{g}\,\left[
{n\over12}R+\mbox{Tr}\,(\hat{E}^\lambda \hat{E}_\lambda)+
\mbox{Tr}\,
(\nabla^\lambda
\hat{E}_\lambda) -\mbox{Tr}\, \left( \hat{\Pi} + {R\over6}
\hat1
\right)
\right], \nonumber
\end{equation}
where the first term comes from the matrix $\hat{\Omega}$
squared.
Calculating these functional traces, we arrive  at
\begin{eqnarray}
\Gamma_{div}=-{1\over2\epsilon}\int d^2x\,\sqrt{g}\,\Biggl\{
\left( {C'' \over Z}  -{n+2m+2\over12}\right)R+{V''\over Z} -
{V_{,ii}\over f}  \qquad\qquad  \cr\cr
+ \left( {{f'}^2\over2fZ}- {f''\over2Z}
\right)(\nabla^\lambda\chi
)(\nabla_\lambda\chi) \qquad\qquad  \cr\cr
+ \left( {mf''\over2f}- {m{f'}^2\over4f^2}
-{{Z'}^2\over4Z^2}\right)(\nabla^\lambda\Phi
)(\nabla_\lambda\Phi)
\qquad\qquad  \cr\cr + \left( {mf'\over2f}- {Z'\over2Z}\right)
\Delta \Phi
+\left({q''\over
         Z}-2{{q'}^2\over qZ}\right)T     \qquad\qquad  \cr\cr
+\left({b''\over Z}-{8b{q'}^2\over qZ}+{16b^2\over
q^2}(N_{ab}N_{ba})\right)
         J^2-{32b^2\over q^2}J(\overline{\psi}N^3\psi)\Biggr\}
\ .
\end{eqnarray}

Expression (44) gives the one-loop effective action of the
system
composed of  quantized dilaton, scalars and Majorana fermions
in an
external gravitational field. It is interesting to note that,
contrary to what happens when the gravitational field itself
is
quantized (Sect. 2), we get a non-trivial renormalization of
the
$C(\Phi)R$-term and $f(\Phi)$ in the kinetic term of scalars.
\bigskip

\section{The one-loop effective action for dilaton gravity
with
Dirac fermions}

In this section we will be interested in the covariant
effective
action of dilaton gravity with Dirac fermions. We will show
that
some technical problems of the covariant formalism may
actually be
solved in this case, after what the calculation will be
performed
precisely as in Sect. 2 for Majorana fermions with dilaton
gravity.

Let us start from the Gross-Neveu model of $n$ Dirac fermions
$\Psi$. The action of such model interacting with dilaton
gravity
is
\begin{equation}
S=\int
d^2x\,\sqrt{g}\,\left[{i\over2}q(\Phi)\overline{\Psi}\gamma^\l
ambda
\rlnabla_\lambda\Psi-b(\Phi)(\overline{\Psi}\Psi)^2\right]+S^{
d.
grav.},
\label{G_N-action}
\end{equation}
where
\begin{equation}
S^{d. grav.}=\int
d^2x\,\sqrt{g}\,\left[{1\over2}Z(\Phi) g^{\mu\nu} \partial_\mu
\Phi
\partial_\nu \Phi+C(\Phi) R + V(\Phi)\right].
\label{G1_N-action} \nonumber
\end{equation}
The functions $b(\Phi)$ and $q(\Phi)$ are supposed to be
smooth
enough.
(Notice that recently a semiclassical approach to the
Gross-Neveu
model
with  Jackiw-Teitelboim dilaton gravity has been discussed in
ref.
[12].)

The equations of motion corresponding to the action yield
\begin{eqnarray}
&
-iq\overline{\Psi}\gamma^\lambda\nabla_\lambda\Psi+{i\over2}
q(\nabla_\lambda J^\lambda)+2bJ^2-2g_{\mu\nu}{\delta
S^{d.grav.}\over\delta
    g_{\mu\nu}}=0 \ , &  \cr\cr
&
-iq\overline{\Psi}\gamma^\lambda\nabla_\lambda\Psi-{i\over2}
q'(\nabla_\lambda\Phi)J^\lambda+2bJ^2=0 \ , &  \cr\cr
&
iq'\overline{\Psi}\gamma^\lambda\nabla_\lambda\Psi
+{i\over2}q''(\nabla_\lambda\Phi)J^\lambda -b'J^2+{\delta
S^{d.grav}\over\delta\Phi}=0 \ . &
\end{eqnarray}
Here we denoted the currents $J=\overline{\Psi}\Psi$ and
$J^\lambda=\overline{\Psi}\gamma^\lambda\Psi$.

To compute the one-loop effective action we find the terms
quadratic in the
quantum fields ($\Psi\to\Psi+\chi\,$, etc.) We write down only
a few  of the most important terms:
\begin{equation}
S^{(2)}={1\over2}\int
d^2x\,\sqrt{g}\,\left[iq\overline{\chi}
\gamma^\lambda\rlnabla_\lambda\chi-iq\bar{h}_{\mu\nu}\overline
{\Psi}
\gamma^\mu\nabla^\nu\chi-2b(\overline{\Psi}\chi)
(\overline{\Psi}
         \chi)+\cdots\right] \ .
\label{non_minimal_terms}
\end{equation}
The last two terms give rise to a non-minimal operator after
the
fermionic
squaring $\overline{\chi}\to\overline{\chi}$, \
$\chi\to-i\gamma^\lambda\nabla_\lambda\chi\,$. One can choose
the
gauge
fixing condition to cancel the second term in
Eq. (\ref{non_minimal_terms}), as
was done in [9], but the last term remains and there is
obviously no
chance of getting rid of it.
This problem always appears in the four-fermi theories but so
far
it
has not been made public since the four-fermi terms are not
renormalizable
on index in the spaces with $d\ge3$.

The problem we tackled never arises in the Majorana case since
with
the help
of the Majorana transposition rules one gets
\begin{equation}
(\overline{\Psi}\chi)(\overline{\Psi}\chi)
=(\overline{\chi}\Psi)
(\overline{\Psi}\chi) \ .
\end{equation}
Hardly could this solve all the problems one encounters in
higher
space-time
dimensions but in $d=2$ it allows to surmount the problem of
the
non-minimality.

Let us introduce two subsets of the Majorana real-valued
fields
$\psi_{1,2}$ according to the rule
$$
\Psi={\psi_1+i\psi_2\over\sqrt2} \ , \qquad
\overline{\Psi}={\overline{\psi_1}-i\overline{\psi_2}
\over\sqrt2}
\ .
$$
The existence of the Majorana representation in two spacetime
dimensions ensures that the fields may be taken
real, so that $\overline{\psi}$ is the {\it charge} conjugate
of  $\psi$, as it should be. Using the Majorana
properties, one has
\begin{eqnarray}
S=\int
d^2x\,\sqrt{g}\,\left[{i\over2}q(\Phi)\overline{\psi_1}
\gamma^\lambda
\nabla_\lambda\psi_1+{i\over2}q(\Phi)\overline{\psi_2}
\gamma^\lambda
\nabla_\lambda\psi_2-{1\over4}b(\Phi)(\overline{\psi_1}\psi_1+
  \overline{\psi_2}\psi_2)^2\right]   \cr\cr
+\mbox{dilaton gravity} \ ,
\end{eqnarray}
where no surface term has been dropped. Notice also that the
covariant derivatives here may be substituted with the
ordinary ones, cf. Eq. (8) above, since the spin connection
drops out of the Majorana bilinears in two dimensions. The
latter
circumstance
make the calculation of the one-loop effective action much
easier
since it
is not necessary to vary the spin connection.

Further, it is profitable to arrange the two Majorana fields
into
a larger
multiplet $\{\psi_a\}_{a=1}^{2n}$ defined as
\begin{equation}
\psi\equiv\pmatrix{\psi_1 \cr \psi_2 \cr} \ .
\end{equation}
Clearly, this is a Majorana field as well, with all its
inherent
properties.
Thus we have
\begin{equation}
S=\int
d^2x\,\sqrt{g}\,\left[{i\over2}q(\Phi)\overline{\psi}
\gamma^\lambda
\partial_\lambda\psi-b(\Phi)(\overline{\psi}N\psi)^2
\right]+\mbox
{dilaton
  gravity}\ ,
\end{equation}
where $N_{ab}={1\over2}\delta_{ab}$.

The scalar current may be recast in the Majorana form as
follows:
\begin{equation}
J\equiv\overline{\Psi}\Psi=\overline{\psi}N\psi \ .
\end{equation}
Also, let us introduce a quantity
\begin{equation}
T^{\mu\nu}\equiv
-{i\over4}\left(\overline{\Psi}\gamma^\mu\rlnabla^{\:\nu}
\Psi+\overline{\Psi}\gamma^\nu\rlnabla^{\:\mu}\Psi\right)
                   \ , \qquad T\equiv
T^\nu_\nu=-{i\over2}\overline{\Psi}\gamma^\lambda
\rlnabla_\lambda
                   \Psi \ ,
\end{equation}
then by the same reasoning as before we get
\begin{equation}
T^{\mu\nu}=-{i\over2}\left(\overline{\psi}\gamma^\mu\nabla^\nu
\psi+
            \overline{\psi}\gamma^\nu\nabla^\mu\psi\right) \ ,
\qquad
T=-i\overline{\psi}\gamma^\lambda\nabla_\lambda\psi \ .
\end{equation}

Now the problem reduces to the old one: that for the Majorana
case,
with only
minor changes. The divergent part of the one-loop effective
action
for the
Gross-Neveu model with the Dirac fields becomes (we use for
this
calculation the results of Sect. 2)
\begin{eqnarray}
\Gamma_{div}=-{1\over2\epsilon}\int
d^2x\,\sqrt{g}\,\Biggl\{{24-n\over6}R
             +{2\over C}V+{2\over C'}V'+\left({C'\over
C}-{Z\over
C'}\right)
             (\Delta\Phi)      \qquad  \cr\cr
+\left({C''\over C}-3{{C'}^2\over
C^2}-{C''Z\over{C'}^2}\right)(\nabla^\lambda
              \Phi
)(\nabla_\lambda\Phi)+\left({3qZ\over4{C'}^2}+{3q\over4C}-
              {q'\over C'}\right)T     \qquad  \cr\cr
+\left({8b^2\over q^2}(n-1)+{4bZ\over{C'}^2}-{4b\over
C}-{2b'\over
C'}-{8bq'
              \over qC'}\right)J^2\Biggr\} \ .
\label{GN_divergences}
\end{eqnarray}

Let us now try to generalize our result to the most general
case of the four-fermi
interaction $(\overline{\Psi}A\Psi)^2$, the matrix $A$ being
an arbitrary
combination of the Dirac algebra elements (some examples are:
$\gamma^\nu\gamma_5$, \
$\gamma^\mu\gamma^\nu\gamma^\lambda\,$,  and so on). The
fact is that all these $A$'s fall into three cases since the
two-dimensional
Dirac algebra basis consists of just three elements: 1,
$\gamma^\mu$ and
$\gamma_5$. Thus the most general situation is
$$
S_{quartic}=-\int
d^2x\,\sqrt{g}\,\Bigl[b_1(\Phi) J^2+ b_2(\Phi)J_5^2+b_3(\Phi)
         J^\mu J_\mu \Bigr] \ ,
$$
with arbitrary smooth functions $b_1(\Phi)$, $b_2(\Phi)$,
$b_3(\Phi)$. The
currents here are defined as usual:
\begin{equation}
J=\overline{\Psi}\Psi \ , \qquad
J_5=\overline{\Psi}\gamma_5\Psi
\ , \qquad
J^\mu=\overline{\Psi}\gamma^\mu\Psi \ .
\end{equation}
However, things may be simplified even further, with the use
of the
two-dimensional Fierz identity which yields:
\begin{equation}
J^\mu J_\mu =J_5^2-J^2 \ .
\label{Fierz_identity}
\end{equation}
Hence we can eliminate one of the three structures and
technically
the
simplest choice is to set $b_3=0$. So, what one has to do is
to
complete the
action (\ref{G_N-action}) with the axial term
\begin{equation}
S_{ax}=-\int
d^2x\,\sqrt{g}\,\Bigl[a(\Phi)(\overline{\Psi}\gamma_5\Psi)^2
         \Bigr] \ .
\end{equation}

Let us turn our attention to Eq. (\ref{Fierz_identity}). At
first glance, it may seem a little strange. The reason is that
its left-hand side is chiral
invariant while both term on the right are not---however the
non-invariant
contributions cancel in pairs. Thus we note that the total
action $S+S_{ax}$ is only chiral invariant if
$a(\Phi)\equiv-b(\Phi)$, and
so are the one-loop divergences. Hence the extra
divergences to arise (cf. Eq.(\ref{GN_divergences}) above) are
\begin{equation}
\Gamma_{ax,div}=-{1\over2\epsilon}\int
d^2x\,\sqrt{g}\,\left({8a^2\over
                 q^2}(1-n)+{4aZ\over{C'}^2}-{4a\over
C}-{2a'\over
C'}-{8aq' \over qC'}\right)J_5^2+\ldots \ ,
\end{equation}
where the dots stand for possible mixed terms proportional to
$ab$. To the end of the section we will compute these terms.

First, go to the Majorana field multiplet $\psi$ so that
\begin{equation}
J_5=i\overline{\psi}\gamma_5M\psi \ , \qquad
M_{ab}=\pmatrix{0 & 1/2 \cr -1/2 & 0 \cr} \ ,
\end{equation}
and
\begin{equation}
J^\mu=i\overline{\psi}\gamma^\mu M\psi \ .
\end{equation}
Note also that
$$
\overline{\psi}M\psi=0 \ ,  \qquad
\overline{\psi}M^2\psi=-{1\over2}J^2 \ .
$$
Second, expand the action
$$
S_{ax}=\int
d^2x\,\sqrt{g}\,a(\Phi)\left(\overline{\psi}\gamma_5M\psi
        \right)^2
$$
in quantum fields $\psi\to\psi+\eta$ and note that the gauge
fixing
terms
(and thus the ghosts) acquire no extra terms. After the
fermion operator
squaring we obtain some {\it additive} corrections to elements
of the matrices $\hat{E}^\lambda$ and $\hat{\Pi}$ (see Eq.
(20)). Therefore, it is the
$\mbox{Tr}\,\big(\hat{E}^\lambda\hat{E}_\lambda\big)$ that
gives the
desired mixed
contribution.

A little thought suggests that the only correction that leads
to $ab$ terms is
\begin{equation}
\delta\big(\hat{E}^\lambda)^4_4=-{2a\over
q}J_5(M\gamma_5\gamma^\lambda)
                                -{4ia\over
q}(\gamma_5M\psi)(\overline{\psi}
                                M\gamma_5\gamma^\lambda)
\end{equation}
so that the desired mixed term contribution is found to be
\begin{equation}
\Gamma_{ax,div}^{mixed\ terms}=-{1\over2\epsilon}\int
d^2x\,\sqrt{g}\:
                         {8ab\over q^2}\left(J_5^2-J^2-J^\mu
J_\mu\right) \ ,
\end{equation}
which is zero by virtue of Eq.(\ref{Fierz_identity}). The
final
answer is
\begin{eqnarray}
\Gamma_{div}=-{1\over2\epsilon}\int
d^2x\,\sqrt{g}\,\Biggl\{{24-n\over6}R
             +{2\over C}V+{2\over C'}V'+\left({C'\over
C}-{Z\over
C'}\right)
             (\Delta\Phi)      \qquad  \cr\cr
+\left({C''\over C}-3{{C'}^2\over
C^2}-{C''Z\over{C'}^2}\right)(\nabla^\lambda
              \Phi
)(\nabla_\lambda\Phi)+\left({3qZ\over4{C'}^2}+{3q\over4C}-
              {q'\over C'}\right)T     \qquad  \cr\cr
+\left({8b^2\over q^2}(n-1)+{4bZ\over{C'}^2}-{4b\over
C}-{2b'\over
C'}-{8bq'
              \over qC'}\right)J^2     \qquad \cr\cr
+\left({8a^2\over q^2}(1-n)+{4aZ\over{C'}^2}-{4a\over
C}-{2a'\over
C'}-{8aq'
              \over qC'}\right)J_5^2 \Biggr\} \ .
\end{eqnarray}

Thus, we have calculated the one-loop divergences of the
covariant
effective action in 2D dilaton gravity within the most general
four-fermionic theory described by Dirac fermions. An
interesting
remark is that the renormalization of all fermionic terms in
the
action is given by the same term (and the same generalized
coupling
constant). For example, if $b (\Phi) =0$ in (45), then the
term
$J^2$ is absent in (65).
\bigskip

\section{The one-loop renormalization}

In the previous sections we have calculated the divergences of
the
one-loop covariant effective action for 2D dilaton gravity
interacting via various kinds of fermionic matter. Let us here
discuss the issue of renormalization, to one-loop order.
Without loss of generality, we will restrict ourselves to the
case of 2D dilaton
gravity with Majorana spinors, viz. Eq. (28).  By adding to
the classical action (8) the corresponding
counterterms ($\Gamma_{div} $  with opposite sign),
one obtains the one-loop renormalized effective action.

Choosing the renormalization of the metric tensor in the
following
form
\begin{equation}
g_{\mu\nu} = \exp\left\{ \frac{1}{\epsilon} \left[
\frac{1}{C(\Phi
)} + \frac{Z(\Phi)}{2{C'}^2 (\Phi)} \right] \right\} \,
g^R_{\mu\nu}
\end{equation}
(this choice absorbs all the divergences of the dilaton
kinetic
term), one can obtain the renormalized effective action
as (for simplicity, we drop the superscript `R'
off
$g_{\mu\nu}$ and $R$):
\begin{eqnarray}
S_R &=& - \int d^2x\,\sqrt{g}\,\left\{ \frac{1}{2} Z
g^{\mu\nu}
\partial_\mu \Phi \partial_\nu \Phi + CR +V +
\frac{1}{\epsilon}
\left( - \frac{V'}{C'} + \frac{ZV}{2{C'}^2} \right) \right.
\nonumber \\
&& + T \left[ q +  \frac{1}{\epsilon} \left(  \frac{q}{8C} -
\frac{Zq}{8{C'}^2}+ \frac{q'}{2C'} \right)\right] \\
&&+ \left. J^2  \left[ b +  \frac{1}{\epsilon} \left(
\frac{3b}{C}
- \frac{3bZ}{4{C'}^2}+\frac{b'}{C'}+ \frac{4bq'}{qC'} -
\frac{8b^2(n-2)}{q^2}\right)\right] \right\}, \nonumber
\end{eqnarray}
where we choose $N_{ab}=\delta_{ab}$.

Now, the conditions of multiplicative renormalizability of the
theory in the usual sense have the following form
\begin{eqnarray}
 - \frac{V'}{C'} + \frac{ZV}{2{C'}^2} &=& a_1V, \nonumber \\
 \frac{q}{8C} - \frac{Zq}{8{C'}^2}+ \frac{q'}{2C'} &=& a_2q,
\\
 \frac{3b}{C} - \frac{3bZ}{4{C'}^2}+\frac{b'}{C'}+
\frac{4bq'}{qC'}
- \frac{8b^2(n-2)}{q^2}  &=& a_3b, \nonumber
\end{eqnarray}
where $a_1$, $a_2$ and $a_3$ are arbitrary constants. These
conditions restrict the form of the functions under
discussion.

Some sets of solutions of Eqs. (68) can be obtained
explicitly. The
simplest choice in the gravity sector is
\begin{equation}
Z=1, \ \ \ \ C(\Phi) =C_1 \Phi, \ \ \ \ V=0,
\end{equation}
where $C_1$ is an arbitrary constant. Choosing also $n=2$, we
get
the following family of renormalizable potentials
\begin{equation}
q(\Phi) = q_1 \Phi^{-1/4} e^{\alpha_1 \Phi},  \ \ \ \ \ \
b(\Phi)
= b_1 \Phi^{-2} e^{\alpha_2 \Phi},
\end{equation}
where $q_1, \ \alpha_1$ and $b_1, \ \alpha_2$ are coupling
constants. The renormalization of these coupling constants
follows
as:
\begin{eqnarray}
q^R_1 &=& q_1 \left[ 1 +  \frac{1}{\epsilon} \left(
\frac{\alpha_1}{2C_1}- \frac{1}{8C_1^2} \right) \right],
\nonumber
\\
b^R_1 &=& b_1 \left[ 1 +  \frac{1}{\epsilon} \left(
\frac{\alpha_2}{C_1}+\frac{4\alpha_1}{C_1}- \frac{3}{4C_1^2}
\right) \right],
\end{eqnarray}
and $\alpha_1$ and $\alpha_2$ do not get renormalized in the
 one-loop approximation. It is interesting to
notice that for $\alpha_1
= 1/(4C_1)$,  $\alpha_2 = -1/(4C_1)$, the coupling constants
$q_1$
and $b_1$ do not get renormalized in the one-loop approach
either.

For $n\geq 2$ these multiplicatively renormalizable potentials
look as

\begin{equation}
q(\Phi)=q_1 \Phi^{-1/4}e^{\alpha_1 \Phi}, \ \ \ \
b(\Phi)=b_1 \Phi^{-3/2}e^{2\alpha_1 \Phi},
\end{equation}
but now $C_1$ is fixed by $(12 a_2-a_3)C_1 + 3/(4C_1)=0$ and
$q_1/b_1 = 16(n-2)C_1$.

Another interesting choice is the following
\begin{equation}
Z= e^{d_1 \Phi}, \ \ \ \ \ C= e^{d_1 \Phi}, \ \ \ \ \ V=0.
\end{equation}
We find in this case (for $n=2$):
\begin{eqnarray}
q &=& q_1 \exp \left[ 2a_2 e^{d_1 \Phi} + \frac{1}{4}
\left(\frac{1}{d_1} -d_1 \right) \Phi \right], \nonumber \\
b &=& b_1 \exp \left[ (a_3-8a_2) e^{d_1 \Phi} -
\left(\frac{1}{4d_1} +2d_1 \right) \Phi \right].
\end{eqnarray}

Now, it is interesting to compare the conditions of
multiplicative
renormalizability (68) with the case when the gravitational
field
is a purely classical one.
Using expression (44) we easily get the renormalized effective
action for the dilaton interacting with Majorana spinors in
curved
spacetime:
\begin{eqnarray}
S_R &=& - \int d^2x\,\sqrt{g}\,\left\{ \frac{1}{2} Z
g^{\mu\nu}
\partial_\mu \Phi \partial_\nu \Phi + CR +V -
\frac{1}{2\epsilon}
\left(  \frac{V''}{Z} + \frac{Z''}{2Z} - \frac{3{Z'}^2}{4Z^2}
\right) \right. \nonumber \\
&& + T \left[ q -  \frac{1}{2\epsilon} \left(  \frac{q''}{Z} -
\frac{2{q'}^2}{qZ} \right)\right] \\
&&+ \left. J^2  \left[ b -  \frac{1}{2\epsilon} \left(
\frac{b''}{Z} - \frac{b{q'}^2}{qZ}+ \frac{16b^2(n-
2)}{q^2}\right)\right] \right\}. \nonumber
\end{eqnarray}
As we see, the conditions of multiplicative renormalizability
look
completely different from (68)
\begin{eqnarray}
\frac{V''}{Z}  = a_1 V, &&  \frac{Z''}{2Z} -
\frac{3{Z'}^2}{4Z^2}
= a_2 Z, \\
  \frac{q''}{Z} - \frac{2{q'}^2}{qZ}   = a_3 q, &&
\frac{b''}{Z}
- \frac{b{q'}^2}{qZ}+ \frac{16b^2(n-2)}{q^2} = a_4b.
\end{eqnarray}

In the same way one can study the renormalization of Dirac
spinorial matter with 2D dilaton gravity. Notice also that one
can
also investigate the renormalization in the $1/n$
approximation;
however, then only the four-fermion term is renormalized.
\bigskip

\section{Conclusions}

In summary, we have studied in this paper the covariant
effective
action approach in 2D quantum dilaton gravity with
four-fermion
models described by Majorana or Dirac spinors. The one-loop
renormalization of the theory has been considered and the
(rather
involved) conditions for multiplicative renormalizability have
been
obtained. The solution of these conditions gives explicit
families
of multiplicatively renormalizable dilaton potentials.These
potentials
maybe the starting points to discuss 2D quantum
dilaton-fermion
cosmology along the ideas expressed in refs [15,16].

One can also investigate the generalized renormalization group
flow
in the models under discussion. To be more specific, let us
consider again the theory (8), and let $T\equiv \{Z,C,q,b,V\}$
be
the set of generalized effective couplings. The general
structure
of the renormalization is now
\begin{equation}
T_0=\mu^{2\epsilon} \left[ T + \sum_{k=1}^\infty \frac{a_{kT}
(Z,C,q,b,V)}{\epsilon^k} \right],
\end{equation}
where, as it follows from (33),
\begin{eqnarray}
a_{1Z} &=& - \frac{Z'}{C'}+\frac{2{C'}^2}{C^2} +
\frac{2ZC''}{{C'}^2}, \nonumber \\
a_{1C} &=& 0 , \nonumber \\
a_{1V} &=& - \frac{V}{C}- \frac{V'}{C'}, \\
a_{1q} &=& - \frac{3qZ}{8{C'}^2}- \frac{3q}{8C} +
\frac{q'}{2C'}, \nonumber \\
a_{1b} &=& - \frac{2bZ}{{C'}^2} + \frac{2b}{C}+
\frac{b'}{C'}+ \frac{4bq'}{qC'} - \frac{8b^2(n-2)}{q^2}.
\nonumber
\end{eqnarray}

Now, the generalized $\beta$-functions can be defined
according to
\begin{equation}
\beta_T=-a_{1T} + Z \frac{\delta a_{1T}}{\delta Z} +C
\frac{\delta
a_{1T}}{\delta C} + V \frac{\delta a_{1T}}{\delta V} +b
\frac{\delta a_{1T}}{\delta b}  +q \frac{\delta a_{1T}}{\delta
q}.
\end{equation}
Applying this rule to the above functions, we get:
\begin{eqnarray}
\beta_Z &=&  \frac{Z'}{C'} + \frac{2{C'}^2}{C^2}-
\frac{ZC''}{{C'}^2}- \frac{2CZ'C''}{{C'}^3} - \frac{4C''}{C}+
\frac{3CZ''}{{C'}^2}, \nonumber \\
\beta_C &=& 0, \nonumber \\
\beta_V &=&  \frac{V}{C} + \frac{V'}{C'}-
\frac{VC''}{{C'}^2}- \frac{CV''}{{C'}^2} +
\frac{2CV'C''}{{C'}^3},  \\
\beta_q &=&  \frac{3q}{8C} -\frac{q'}{2C'} -
\frac{3CZq'}{4{C'}^3}-
\frac{3CqZ'}{4{C'}^3}+ \frac{9CqZC''}{{C'}^4} +
\frac{Cq''}{2{C'}^2}
\nonumber \\ && -\frac{Cq'C''}{{C'}^3}- \frac{3qZ}{8{C'}^2} +
\frac{qC''}{2{C'}^2} , \nonumber \\
\beta_b &=& -\frac{2b}{C} -\frac{5b'}{C'} -
\frac{2bZ}{{C'}^2}+
\frac{5bC''}{{C'}^2}+ \frac{Cb''}{{C'}^2} +
\frac{8(n-2)b^2}{q^2}
\nonumber \\ && -\frac{2Cb'C''}{{C'}^3}- \frac{4Cb'Z}{{C'}^3}
-
\frac{4CbZ'}{{C'}^3} +\frac{12CbZC''}{{C'}^4}  \nonumber \\
 && +\frac{4Cb'q'}{q{C'}^2}+ \frac{4Cbq''}{q{C'}^2} -
\frac{4Cb{q'}^2}{q^2{C'}^2} -\frac{8Cbq'C''}{q{C'}^3}.
\end{eqnarray}

The renormalization group fixed points of the system under
discussion are defined by the zeros of the above
$\beta$-functions
(see also [10]). What we obtain is the following. As is not
difficult
 to see, the same structure of fixed points that we analyzed
in
detail in our  paper [7] is maintained here. In fact, the
first three of the beta functions are exactly the same as the
ones
for that restricted case, and it suffices to impose $q (\Phi)
\equiv
0$  and  $b (\Phi) \equiv 0$ in order to obtain corresponding
fixed points in the present, generalized case. However, one
has to
take care of the limits, that is, actually we must put $q
(\Phi) =
\eta =$ const., where $\eta$ is arbitrarily small, in order
that
 the families of fixed points obtained in [7] give also
corresponding families here, which are approached as $\eta
\rightarrow 0$. We shall not repeat this construction
 here and simply refer the reader to this paper.

In the same way, the case of the more general four-fermion
theory can
also be discussed.

\vspace{5mm}

\noindent{\large \bf Acknowledgments}

S.D.O. wishes to thank H. Osborn and A. Tseytlin for useful
discussions on related problems,
T.Banks for helpful remark, the Japan Society for the
Promotion of
Science (JSPS, Japan) for financial support  and the
Particle Physics Group at Hiroshima University for kind
hospitality.
E.~Elizalde is grateful to Prof.~I.~Brevik and
Prof.~K.~Olaussen,
and to
Prof.~L.~Brink for the hospitality extended to him at the
Universities
of Trondheim and G\"{o}teborg, respectively.
This work has been  supported by DGICYT (Spain) and by CIRIT
(Generalitat de Catalunya).
\bigskip \bigskip

\appendix

\section{Appendix}

In this short appendix we shall see how our results give the
one-loop renormalization of such a well-known model as the
sine-Gordon one.

Let us start from the action
\begin{eqnarray}
S &=&  - \int d^2x\,\sqrt{g}\,\left[ \frac{1}{2}  g^{\mu\nu}
\partial_\mu \Phi \partial_\nu \Phi + \sqrt{2} \Phi R + \mu \,
e^{-
\sqrt{2} \Phi} \right. \nonumber \\
&&+ \left. m \cos (p\chi )e^{\alpha\Phi} +   \frac{1}{2}
g^{\mu\nu} \partial_\mu \chi \partial_\nu \chi \right].
\end{eqnarray}
Here $p$ is some number and $m$, $\alpha$ and  $\mu$ are
coupling
constants. The one-loop effective action of this theory has
been
calculated in Sect. 2, Eq. (28). Using this result and making
the
renormalization of the metric according to
\begin{equation}
g_{\mu\nu}= \exp \left[\frac{1}{\epsilon\sqrt{2}\Phi} \right]
\,
g_{\mu\nu}^R,
\end{equation}
one obtains the renormalization of the coupling constants in
the
following way
\begin{equation}
\mu_R = \mu \left( 1+ \frac{1}{\epsilon} \right), \ \ \ \ \
m_R = m \left[ 1+ \frac{1}{\epsilon} \left( \frac{p^2}{2} -
\frac{\alpha}{\sqrt{2}} \right) \right].
\end{equation}
The coupling constant $\alpha$ is not renormalized at one-loop
order. Hence, there is no interesting renormalization group
dynamics at one-loop approximation. (It is well known that, in
fact,
interesting dynamics appear in the non-perturbative approach,
as in
the matrix models [13]).


\newpage



\begin{thebibliography}{99}

\bibitem{} C.G. Callan, S.B. Giddings, J.A. Harvey and A.
Strominger,  Phys. Rev. {\bf D45} (1992) 1005;
E. Witten, Phys.Rev. {\bf D44} (1991) 314;
J.A. Harvey and A. Strominger, preprint EFI-92-41
(1992);
L.Susskind ,L.Thorlacius and J.Uglum ,hep-th 9306069.

\bibitem{} S. de Alwis, Phys. Rev. {\bf D46} (1992) 5438; K.
Hamada
and A. Tsuchiya, preprint UT-Komaba 92-14 (1992); A. Bilal
and C. Callan, Nucl. Phys. {\bf B 394} (1993) 73;
 S. Nojiri and I. Oda, Mod. Phys. Lett. {\bf A8} (1993) 53;
 S. Hirano, Y. Kazama and Y. Satoh,
preprint UT-Komaba 93-3;  S. Giddings and A.
Strominger, preprint UCSB-TH-92-28 (1992);
Y. Tanii, Phys. Lett. {\bf B302} (1993) 191.

\bibitem{} S.D. Odintsov and I.L. Shapiro, Phys. Lett. {\bf
B263}
(1991) 183; Mod. Phys. Lett. {\bf A7} (1992) 437; Int. J. Mod.
Phys. D, to appear.

\bibitem{} T. Banks and M. O'Loughlin, Nucl. Phys. {\bf B362}
(1991) 649; preprint RU-92-61 (1992).

\bibitem{} R. Kantowski and C. Marzban, Phys. Rev.
{\bf D46} (1992) 5449; E. Elizalde and S.D. Odintsov, Nucl.
Phys.
B, to appear;  E. Elizalde, S. Naftulin and S.D. Odintsov,
Yad.
Fiz. (Sov. J. Nucl. Phys.), to appear;  F.D. Mazzitelli and N.
Mohammedi,  Nucl. Phys. B, to appear.

\bibitem{} J.G. Russo and A. Tseytlin, Nucl. Phys. {\bf B382}
(1992) 259.

\bibitem{}  E. Elizalde, S. Naftulin and S.D. Odintsov,
preprint
HUPD93-10 (1993), Phys. Lett. B, to appear.

\bibitem{} R. Jackiw, Nucl. Phys. {\bf B252} (1985) 343;
 C. Teitelboim, Phys. Lett. {\bf B126} (1983) 41;
 R. Mann, in Proc. 4th
Canadian Conf. on GR and RA, to be published; I.M. Lichtzier
and
S.D. Odintsov, Mod. Phys. Lett. {\bf A6} (1991) 1953; D.
Cangemi
and R. Jackiw, Phys. Rev. Lett. {\bf 69} (1992) 233; Ann.
Phys., to
appear; G. Crignani and. G. Nardelli, preprint UTF-292-1993;
A.
Chamseddine, Phys. Lett. {\bf B256} (1991) 379.

\bibitem{} A.O. Barvinski and G.A. Vilkovisky, Nucl. Phys.
{\bf
B191} (1981) 237.
\bibitem{} G. 't Hooft and M. Veltman, Ann. Inst. H. Poincare
{\bf
20} (1974) 69.

\bibitem{} I.G. Avramidi, preprint Rostov State Univ. (1993);
Phys.
Lett. {\bf B305} (1993) 27.

\bibitem{} T. Muta and S.D. Odintsov, Progr. Theor. Phys., to
appear.

\bibitem{} G.Moore, hep-th 9203061;
E.Hsu and D.Kutasov ,Nucl Phys. {\bf B396} (1993) 693.

\bibitem{} I.L.Buchbinder, S.D.Odintsov and I.L.Shapiro,
Effective action in quantum gravity ,IOP publishing,
Bristol and Philadelphia, 1992.

\bibitem{} J.Polchinski, Nucl.Phys.{\bf B394} (1989) 123.

\bibitem{} A.Cooper, L.Susskind and L.Thorlacius,
Nucl.Phys. {\bf B263} (1991) 132.

 \end{thebibliography}
\end{document}